# Evaluation of Peak Shaving using Thermal Energy Storage in a Validated CHP and District Energy Model

**Michael Huylo, P.E.**  **Sina Taheri, Ph.D.**  **Atila Novoselac, Ph.D.**
*Member ASHRAE*  *Member ASHRAE*

**ABSTRACT**

*There is currently a large federal effort to decarbonize the country's electrical grid as part of the clean energy transition. The elimination of fossil fuel fired systems, and their replacement with intermittent renewable sources and other electric equipment will require better load management techniques to ensure a reliable grid. One strategy for maintaining electric grid reliability utilizes peak shaving. Buildings, accounting for 40% of energy use in the United States, can account for an even higher percentage of energy during peak periods driven by high air conditioning loads during the summer, especially in hotter climes such as Austin, Texas. Many previous studies have modeled the effectiveness of building HVAC demand response methods such as temperature setpoint manipulation, pre-cooling, ventilation scheduling, and thermal energy storage. Thermal storage systems, due to their larger energy capacities, have been shown to be most promising for peak shaving. However, there is a lack of work integrating chilled water energy storage models with validated microgrid-district energy system models to fully capture the dynamics of the proposed strategies.*

*Previously, a validated system model for power generation and heating was developed for the University of Texas at Austin (UT Austin). A new validated model integrates the 65 MW combined heat and power plant (CHP), with the campus' 45,000 ton district cooling system, as well as two chilled water storage tanks. While the existing campus system currently utilizes an operator driven peak shaving strategy utilizing thermal storage, optimization results show that there is room for further improvement and energy savings. The presented results quantify the peak shaving in MW and provide a foundation for further analysis.*

**INTRODUCTION**

Peak load shaving is an important strategy for utility operators to maintain grid reliability. As energy use continues to rise and global climate change increases peak loads driven by air conditioning, a great deal of stress is placed on existing grid infrastructure. Peak loads make up a very small percentage of overall electricity demand but have an outsized effect on equipment sizing and utility operation. Typically, supplementary "peaking" generators that are less efficient and produce more carbon emissions are needed to generate excess loads required during peaks. Peak load shaving in large utility grids is a well-studied subject providing benefits for both grid operators and end users. Shaving peak loads can improve power quality, increase generation efficiency, reduce costs, reduce carbon emissions, reduce transmission and distribution losses and improve grid reliability (Uddin et al. 2018). Energy storage systems and demand response strategies are common methods used to shift and shave peak loads. Peak shaving in microgrids is less studied but is still an important method in maintaining a reliable system, especially in islanded (isolated from larger utility grid) type microgrids. The benefits may be less obvious, or

**Michael Huylo** is a PhD student in the Department of Civil, Architectural, and Environmental Engineering at the University of Texas at Austin, Austin, Texas. **Sina Taheri** is a machine learning and optimization engineer at Ibotta, Inc. Denver, Colorado. **Atila Novoselac** holds the Marion E. Forsman Centennial Professorship in Engineering in the Department of Civil, Architectural, and Environmental Engineering at the University of Texas at Austin, Austin, Texas.

more difficult to quantify than in a typical grid system where varying time of use rates can better motivate end users to shift or shave their own loads to save money when purchasing electricity. In the case of a microgrid operator who is generating their own electricity from more stable fossil fuel prices such as natural gas, there may appear to be less incentives to reducing peak loads. However, improved power quality, more efficient energy utilization, reduced distribution losses, and easier integration of distributed energy resources (DER) have been shown as benefits in microgrid peak shaving (Rana, Atef, et al. 2022). In many large campus systems, such as the University of Texas at Austin (UT Austin), microgrids are utilized along with district cooling systems and thermal energy storage, the latter of which can provide peak shaving by reducing chiller electricity demand during hot summer afternoons.

One of the review studies on peak shaving in microgrid systems (Rana, Atef, et al. 2022) describes many of the benefits previously mentioned here as well as outlines the need for further work in the optimal scheduling of energy storage systems to adequately reduce microgrid peak loads; authors also used real data to model a hybrid photovoltaic-battery type system to provide peak shaving in a real microgrid system using MATLAB (Rana, Rahman, et al. 2022). Two other studies (Uddin et al. 2020; Ganesh, Senthilmurugan, and Ananthanarayanan 2023) both used MATLAB to model peak shaving in an islanded microgrid system using hypothetical battery energy storage. In another study, Sigrist and Zhang developed optimization algorithms using hypothetical battery systems for their microgrid systems while also presenting economic savings results (Sigrist, Lobato, and Rouco 2013; Zhang et al. 2021). Furthermore, Saboori presented results quantifying the reduction in distribution losses from using energy storage for peak shaving (Saboori and Abdi 2013); Chapaloglou modeled both thermal and electrical systems in a microgrid but utilized the surrounding electrical grid for supplementary power and operated a heat pump system (Chapaloglou et al. 2021); Cole, Ooka, Kamal, and Schutz published works studying the optimal use of thermal energy storage but all did not incorporate islanded microgrid systems with real data or district cooling systems with real data (Cole, Powell, and Edgar 2012; Ooka and Ikeda 2015; Kamal et al. 2019; Schütz, Streblow, and Müller 2015). All these studies provided valuable insight, but there is currently a lack of literature in integrated modelling of islanded microgrid systems including combined heat and power (CHP) plants, district cooling/chiller plants, and thermal energy storage systems all based on measured data. The campus at UT Austin is uniquely suited for such modelling as it utilizes all these system types and currently uses an operator decision strategy for shaving peak microgrid loads with chilled water storage tanks. The development of an integrated CHP, district cooling, and thermal storage model can be used to identify potential peak load reductions using an optimization algorithm.

As a nonprofit academic institution with a large CHP system, UT Austin is not participating in the electricity market, and therefore it is unable to sell excess electricity to the surrounding grid. At the same time, the UT system does not purchase electricity from the surrounding grid (as it produces it for a fraction of the market price) unless in the case of emergencies. Most often, the UT campus is powered by one set of gas and steam turbine generators. During peak loads, a supplementary peaking steam turbine is used to provide the excess power required; however, this system operates at a lower efficiency in generating electricity and requires steam generation from a boiler as there is no longer excess waste heat from the gas turbine combustion gases to generate adequate steam at these loadings. Further, as peak loads approach the maximum design capacity of the microgrid, distribution losses increase with the square of the total current. While distribution losses will not be modeled as part of this study, it is mentioned as a motivation for providing peak shaving in a microgrid system.

Previously, a validated CHP model of the UT Austin system was developed for the incorporation of hypothetical renewable energy generation (Huylo and Novoselac 2024). In this present study, the existing model, with renewable generation aspects removed, was integrated with a validated model of the campus' chilled water district cooling system and chilled water storage tanks. The simulation results from the developed optimization algorithm provide peak shaving results improving on the current system's strategy.

## METHODOLOGY

UT Austin utilizes a CHP plant that produces all the campus' electrical and heating needs from two gas turbine generators, two large steam turbine generators, two smaller supplementary steam turbine generators, four steam boilers, and two heat recovery steam generators. In typical operation one of each component will run together, while the others are off, in standby, or undergoing maintenance. One group of equipment is used for the warmer months while the other is used for winter and shoulder season. In the previously mentioned study one grouping of these components was modeled from operating data. These components are presented in Figure 1, and include a gas turbine generator (CTG 10) with nominal capacity of 32 MW of electrical power (~110 MBTU), a heat recovery steam generator (HRSG 10) with nominal capacity of 50 MW of heat (~170 MBTU), a steam boiler (B7) with nominal capacity of 88 MW of heat (~300 MBTU), and a steam turbine generator (STG 7) with nominal capacity of 25 MW of electrical power (~85 MBTU). In this new model, an additional component is considered, the smaller 8 MW (~17 MBTU) supplementary peaking steam turbine generator (STG). Figure 1 displays the connections between the different pieces of equipment.

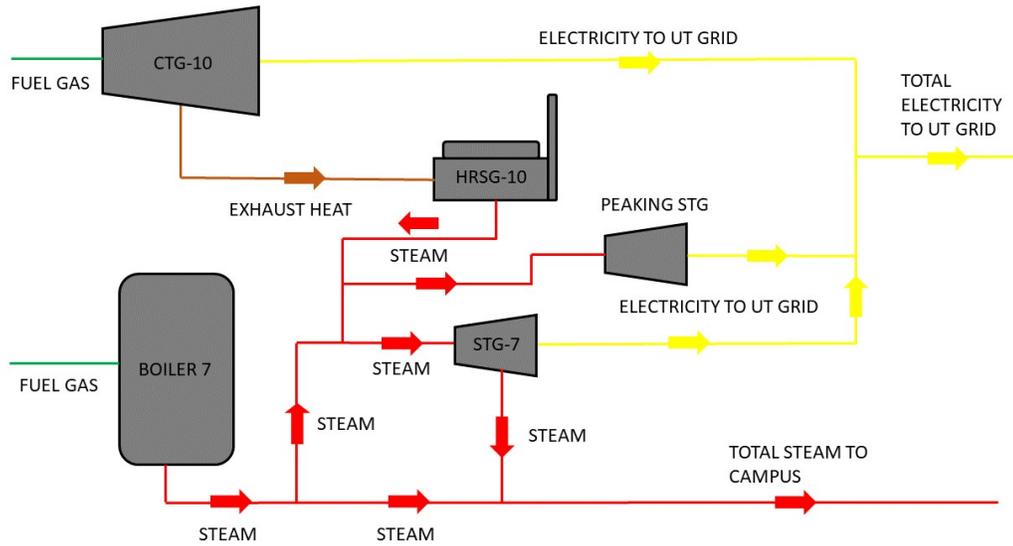

**Figure 1**  A schematic of the UT Austin CHP components displaying inputs and outputs.

Regression models for the components were developed in the previous study and a set of equations was constructed by conducting a heat and mass balance using the Figure 1 schematic. The set of equations were then used as constraints for the optimization problem. Additional constraints for equipment upper and lower bounds were also included.

**Table 1. CHP System Constraints**

| Constraint # | Constraint |
|---|---|
| 1 | $P\_E\_C = P\_E\_ST + P\_E\_GT$ |
| 2 | $Q\_S\_C = Q\_EX\_ST + Q\_S\_SB \ast PRR$ |
| 3 | $Q\_S\_SB = Q\_S\_SB \ast PRR + Q\_SB\_ST$ |
| 4 | $Q\_S\_ST = Q\_S\_HRSG + Q\_SB\_ST$ |
| 5 | $P\_E\_GT = Q\_G\_GT \ast \eta GT$ |
| 6 | $Q\_HEAT = Q\_G\_GT - P\_E\_GT$ |
| 7 | $Q\_S\_HRSG = Q\_HEAT \ast \eta HRSG$ |
| 8 | $Q\_S\_SB = Q\_G\_SB \ast \eta SB$ |
| 9 | $Q\_EX\_ST = Q\_S\_ST \ast ER$ |
| 10 | $P\_E\_ST = Q\_S\_ST \ast \eta ST$ |

Here, P_E_C, P_E_ST, and P_E_GT are the campus electrical load, electricity generated by the steam turbine, and electricity generated by the gas turbine respectively. Q_S_C is the campus steam load. Q_EX_ST is the steam extracted from the steam turbine and Q_S_SB is the steam generated at the steam boiler. PRR is the "pressure reducing ratio", or fraction of steam from the steam boiler that goes directly to campus. Q_SB_ST is the steam from the steam boiler directed to the steam turbine. Q_S_ST is the total steam entering the steam turbine and Q_S_HRSG is the steam generated at the HRSG. Q_G_GT and Q_G_SB are the fuel gas burned at the gas turbine and steam boiler respectively. Q_HEAT is the amount of waste heat from gas turbine exhaust gases entering the HRSG, and ER is the "extraction ratio", or ratio of steam entering the steam turbine that is extracted and sent to campus to provide some or all of the required heating load. Even during the summer there is still some steam demand for process uses or domestic water heating. The variables ηGT, ηHRSG, ηSB, and ηST are the component regression models. In the case where system electrical loads exceed the capacity of the primary components CTG-10 (32MW) and STG-7 (25MW) for a total of 57 MW, the remaining load must be provided by the peaking STG. The peaking STG is older, and less efficient at generating electricity. Additionally, when the CHP system reaches the 57 MW threshold, it has already fully utilized all of the waste heat from the gas turbine generator. Therefore, any electricity provided by the peaking STG must be produced by steam from boiler 7 using additional fuel. See the previously mentioned study for more details on the development of the CHP model if required (Huylo and Novoselac 2024).

For this work, a model of the UT Austin district cooling system was developed from measured operating data to provide additional constraints for the peak shaving optimization problem. A schematic of the campus cooling system is shown in Figure 2. The campus currently has five different chiller stations (two older stations have been replaced) each with different chillers and system configurations. Chiller station 6 is the newest and most efficient station and usually carries a significant portion of the campus load. Chiller station 4 is older and less efficient and is used less often (if other stations are inoperable due to maintenance). The campus includes two thermal energy storage tanks where generated chilled water is stored overnight to help reduce peak loads the following afternoon.

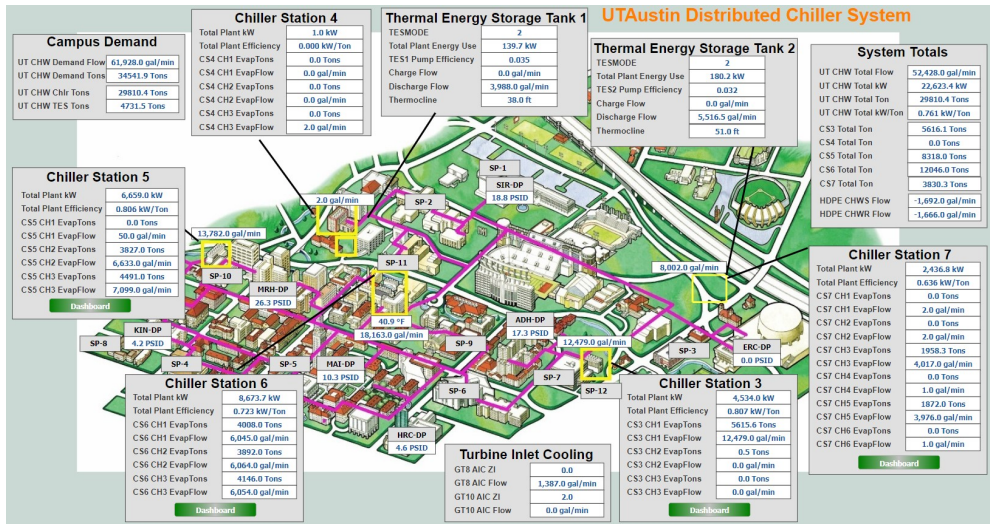

**Figure 2** A schematic of the UT Austin district cooling system including chiller stations and thermal energy storage.

Constraints for the coefficient of performance (COP), total chilled water storage tank volume, discharging and charging rate of the chilled water storage, current state of charge, state of charge at beginning and end of simulation period, and the contributions to the delivered cooling load from both storage and chillers at each time step were developed from available data. The two chilled water storage tanks were lumped into one larger tank for the purposes of the model. The cooling system model was coupled to the CHP model through constraint #1 in Table 1. Previously, the total electrical power draw of the campus P_E_C was a measured value and was set equal to the sum of the electricity provided by the steam and gas turbine generators. In the new model the left side of constraint #1 becomes the sum of the base load required on campus, P_Base (which is a real measured value that represents all electricity demand except that from the district chilled water cooling system) and P_CH, which is the power required by the chilled water system which will be optimized over 24-hour intervals to flatten the peak load on sample test days. The constraints for the chilled water system are listed in Table 2.

In Table 2, PLR is the partial load ratio, and TWB is the wet bulb temperature in Celsius. The nominal cooling of the entire system is 45,000 tons or 156.5 MW. Q_CH is the cooling generated by the chiller system, P_CH is the electrical power draw from the chiller system, Q_COOL is the overall cooling delivered to campus, E_STOR is the thermal energy stored in the chilled water tank which has a maximum value of 175.6 MWh or approximately 50,000 ton-hrs, and Q_STOR is the thermal energy storage charging or discharging rate. The thermal energy stored is set to the maximum value at both the beginning of each day, and in the last hour of each day. The thermal energy stored at the next time step is equal to the previous time step's storage plus the charging/discharging rate. The rate of charge/discharge is limited to 31.7MW per hour or 9,000 tons.

**Table 2. Chilled Water-Cooling System Constraints**

| Constraint # | Constraint |
|---|---|
| 11 | COP = 11.87 - 8.84*PLR - 0.17*TWB - 6.89*PLR$^2$ + 0.75*TWB*PLR - 0.01*TWB$^2$ |
| 12 | PLR = Q_CH/156.5 |
| 13 | P_CH*COP = Q_CH |
| 14 | Q_COOL = Q_CH - Q_STOR |
| 15 | E_STOR(1) = 175.6 |
| 16 | E_STOR(T+1) = 175.6 |
| 17 | E_STOR(i+1) = E_STOR(i) + Q_STOR |
| 18 | -31.7 <= Q_STOR <= 31.7 |

To accurately predict the power consumed by the campus district cooling system a reduced order regression model considering the different chiller stations was developed in scikit-learn. One year of hourly data for wet bulb temperature, cooling generated, and electricity consumed (for a lumped model of all five of the campus chilling stations) was used to generate the multi variable regression equation for the campus chilling system COP. The model predicts system COP as a function of both wet bulb temperature and partial load ratio. Typically, the operators maintain the same staging sequence of stations and individual chillers depending on load, so a reduced order multi variable regression model was able to accurately capture the combined power draw from each station depending on overall loading. The wet bulb temperature adjusts the system's performance depending on the cooling tower's ability to reject heat. The regression model was constructed from overall system power ensuring that power requirements for chilled water pumps, condenser water pumps, district loop distribution pumps, and condenser fans were also included. The error metrics for the coefficient of the variation of the root mean square error (CVRMSE) and the mean bias error (MBE) were calculated according to ASHRAE Standard 14 guidelines (ASHRAE 2014). The validation results for the COP model satisfied the ASHRAE 14 simulation requirements and are displayed in Table 3. Note that the MBE is a positive number, meaning the model slightly underestimates COP, which means the peak shaving results calculated later may be conservative.

**Table 3. Chiller System COP Model Validation Results**

| Test Day | CVRMSE | MBE |
|---|---|---|
| June 1-4 2023 | 2.6% | 0.4% |
| September 10-13 2023 | 4.3% | 2.5% |

Last, the objective function to be minimized for this problem to shave peak loads is equal to (P_E_GT + P_E_ST – P_Mean)². The sum of the campus' electrical load is produced from total gas turbine power and total steam turbine power i.e. the sum of these productions (P_E_GT + P_E_ST). P_Mean is the average (or completely flattened) electrical demand from the previous day, where it is assumed weather and operating conditions are relatively similar. Therefore, (P_E_GT + P_E_ST – P_Mean)² as an objective function represents the linear least squares from a completely flat load profile (ideal to reduce peaks), which forces the system to reduce demand during peaks, and accommodate for reduced peaks by using more energy in "valleys". In order to reduce peaks, the electrical consumption of the system must go down during hours of peak demand. The base electrical load consumed by campus in our problem remains the same, the variable that can be optimized to reduce the electrical peak is the overall power draw from the chiller system or P_CH. The chiller system reduces its electrical consumption during peak hours by discharging/charging the thermal storage in an optimal way over the 24-hour period. The coupled constraints between systems (CHP and chilled water) drive the optimal charging/discharging of the thermal storage tank.

The optimization code was written in MATLAB using the fmincon function and the interior point algorithm from the optimization toolbox. Max iterations were set to 1e5 and max function evaluations were set to 1e5. The calculations were performed using MATLAB R2024a on a windows operating system with 8.00 GB of RAM and an AMD Ryzen 5 2600 Six-Core Processor 3.40 GHz. Optimization results for one 24-hour period took approximately 15-20 seconds. For larger simulations multiple days can be simulated at the same time using the MATLAB parallel computing toolbox.

**RESULTS**

**Optimal Load Shaving**

Two 72-hour periods from the summer were chosen to evaluate the optimization model's ability to shave peak loads using the integrated CHP and district cooling/thermal energy storage model. Plots are presented for the real measured hourly peak loads (baseline) as well as the optimal shaving results from the model. Note that the UT Austin system is already performing peak shaving using the thermal energy storage system, so there is less room available for improvement, but it is desirable to see if the current operator manually controlled method (baseline) is close to optimal. The exact current operator-driven control inputs are not available to us, we only have their results as a baseline to compare our optimization method. It is our understanding that depending on which particular operator is working at any time, they may operate the charging and discharging of the thermal storage slightly differently based on their own rule of thumb, but typically charging begins in the late evening (~8-11 PM) and completes in the early morning (~3-6 AM), and it is then discharged during the afternoon. Additional calculations were also conducted to estimate what the peak load on the UT system would be if there were no storage system or peak shaving strategy in place. The first presented results were from September of 2023 where outdoor dry bulb temperatures reached 100°F (38°C) and buildings were fully occupied as the academic year had started. The results for September 2023 are displayed in Figure 3.

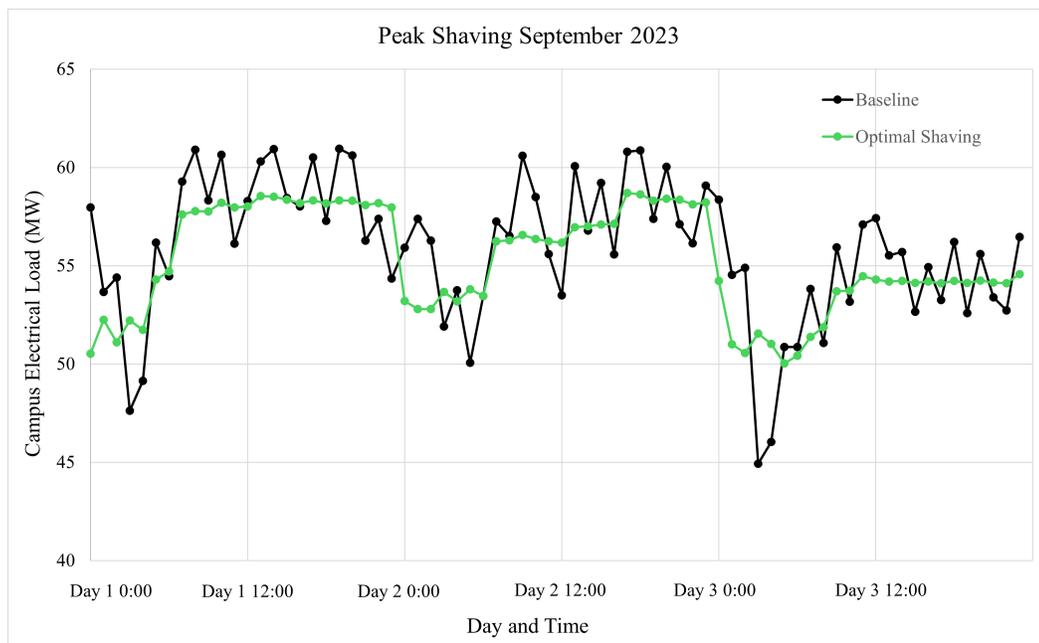

**Figure 3** Results for the optimal peak load shaving for a 72-hour period in September 2023.

First, note that in the estimated calculations it was determined that peaks during the first and second day would have reached approximately ~65MW if not for the existing thermal storage system already providing some peak reduction. However, even having less room for improvement by optimizing the performance of the integrated system, the optimization algorithm was regularly able to reduce peaks by 3-5 MW (~5-7%). Second, the optimized results present a flatter curve with significantly reduced oscillations, allowing some periods of higher energy use to offset sharp peaks. Third, as the sum of the two main generators is equal to 57MW, the peak load shaving was almost able to eliminate the use of the more costly supplementary peaking steam turbine on the first two days and was able to completely eliminate it on the third day. Even if the first two days were reduced by another 1 or 2 MW, it would still require the smaller steam turbine to remain on in a minimum position for added system reliability in the case of a slight load increase. Additional savings would also be realized from the reduction in distribution losses. Similar results for another 72-hour period, this time from June of 2023, are presented in Figure 4. Here, the first day has an even higher loading than presented in Figure 3.

In Figure 4 the maximum peak load is slightly more than 64 MW during the first afternoon. Our model calculations estimated that the peak load during this period would have been ~67-68 MW if there was no thermal storage system present on campus. Again, the optimization model was able to outperform the current operator driven peak shaving method, reducing peak loads by as much as 5 MW (5-8%). In the first 24-hour period the use of the peaking steam turbine was nearly eliminated, and it was entirely eliminated on the second and third day. Note that the second day did not cross the 57 MW threshold but was at times close enough that it would have been activated at a minimum loading for reliability purposes. It is also worth noting that in both test periods the elimination or reduction in use of the peaking steam turbine will also reduce carbon emissions, as the flatter generation curves can better utilize gas turbine waste heat while reducing supplementary steam boiler use.

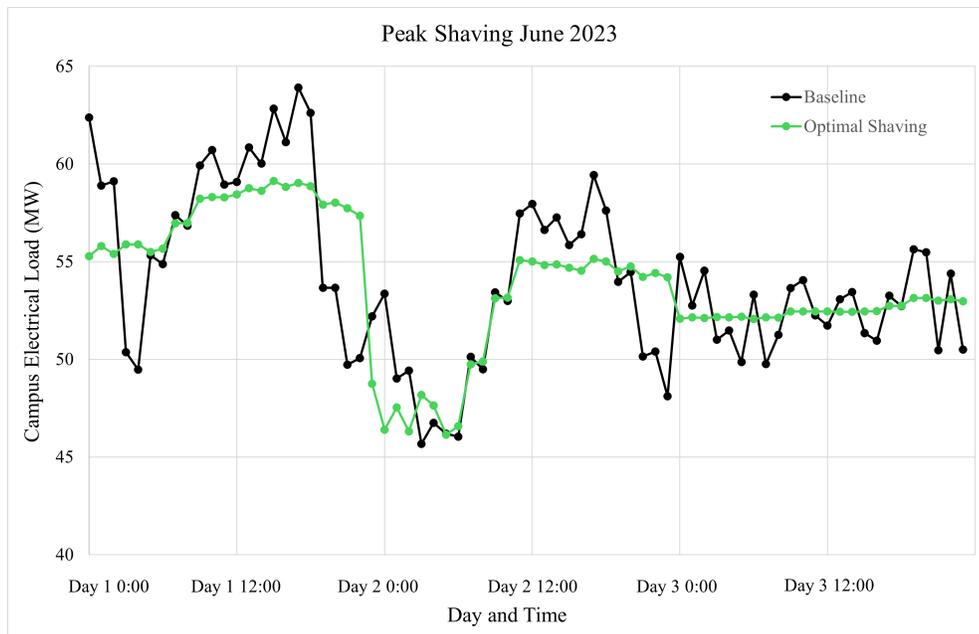

**Figure 4**   Results for the optimal peak load shaving for a 72-hour period in June 2023.

Last, knowing the electrical generation efficiency of the combined cycle gas turbine and steam turbine, and knowing the electrical generation efficiency of the peaking steam turbine, the peak shaving fuel savings can be roughly calculated. Assuming that above the 57 MW threshold electricity is generated by the peaking turbine, and below the 57 MW threshold is generated by the combined cycle, the area between the two curves above the 57 MW threshold will provide fuel savings. In reality this would not be entirely accurate due to the fact that the peaking turbine would need to be staged on at a minimum load as the demand approached this threshold, however it can provide a reasonable approximation. For example, in mid-morning to late afternoon on day 1 in Figure 3, the load is shaved from approximately 61 MW to 58 MW. This results in a peak reduction of about 5% but in these particular hours the fuel savings were almost 9%. In another example, at 5 PM on day 1 in Figure 4, the load is shaved from approximately 64 MW to 59 MW. This was a peak reduction of about 8% but a fuel reduction of almost 14%. However, these larger saving opportunities were only available in hours when the peaking turbine was required. Across both 72-hour periods in Figure 3 and Figure 4 the total cumulative fuel savings was approximately 1%. These calculations also did not consider the savings available from the reduction in distribution losses.

## CONCLUSION

A previously developed system model of the combined heat and power plant at the University of Texas at Austin was combined with a new model of the campus district cooling system and thermal energy storage tanks. Real data was used to generate a reduced order model that accurately predicts the campus chilling system coefficient of performance. Thermal energy storage capacities and charge/discharge rates were also defined. The two models were coupled through the campus power balance constraint. The electricity from the steam and gas turbines was set equal to a base electrical load taken from measured data, and a variable for the electrical draw for the chiller system. The system was optimized to minimize the sum of the square of the difference between the generator powers and a mean hourly power over a period of 24 hours. It was found that the developed optimization algorithm could consistently reduce peak afternoon loads by 5% or more leading to savings in generation costs and distribution costs. Further work will involve developing a resistance model of the campus microgrid distribution system to quantify the savings from reduced resistive distribution losses. An expanded study will also consider more peak events over the course of the summer.

## ACKNOWLEDGEMENTS

This paper is in part, based upon work supported by the U.S. Department of Energy's Office of Energy Efficiency and Renewable Energy (EERE), Award Number DE-EE0009139. This paper is also in part based upon work supported by the 2024-2025 ASHRAE Graduate Student Grant-In-Aid award.

## REFERENCES

ASHRAE. 2014. *ASHRAE Guideline 14: Measurement of Energy and Demand Savings.* Atlanta, GA: American Society of Heating, Refrigerating, and Air-Conditioning Engineers Inc.

Chapaloglou, Spyridon, Athanasios Nesiadis, Konstantinos Atsonios, Nikos Nikolopoulos, Panagiotis Grammelis, Angel Carrera, and Oscar Camara. 2021. "Microgrid energy management strategies assessment through coupled thermal-electric considerations." *Energy Conversion and Management* 228: 113711. https://doi.org/https://doi.org/10.1016/j.enconman.2020.113711. https://www.sciencedirect.com/science/article/pii/S0196890420312358.

Cole, Wesley J., Kody M. Powell, and Thomas F. Edgar. 2012. "Optimization and advanced control of thermal energy storage systems." 28 (2-3): 81-99. https://doi.org/doi:10.1515/revce-2011-0018. https://doi.org/10.1515/revce-2011-0018.

Ganesh, Viswanathan, S. Senthilmurugan, and Rathinam Ananthanarayanan. 2023. "Implementation of Peak Shaving Algorithm in an Islanded Microgrid for Economic Power Consumption." Smart Energy and Advancement in Power Technologies, Singapore, 2023//.

Huylo, Michael, and Atila Novoselac. 2024. "CHP Digital-Twin-Model for Control, Optimization and Transition to Renewable Based Fuels." *ASHRAE Transactions* 130 (1): 121-129.

Kamal, Rajeev, Francesca Moloney, Chatura Wickramaratne, Arunkumar Narasimhan, and D. Y. Goswami. 2019. "Strategic control and cost optimization of thermal energy storage in buildings using EnergyPlus." *Applied Energy* 246: 77-90. https://doi.org/https://doi.org/10.1016/j.apenergy.2019.04.017. https://www.sciencedirect.com/science/article/pii/S0306261919306427.

Ooka, Ryozo, and Shintaro Ikeda. 2015. "A review on optimization techniques for active thermal energy storage control." *Energy and Buildings* 106: 225-233. https://doi.org/https://doi.org/10.1016/j.enbuild.2015.07.031. https://www.sciencedirect.com/science/article/pii/S0378778815301420.

Rana, Md M., Mohamed Atef, Md R. Sarkar, Moslem Uddin, and G. M. Shafiullah. 2022. A Review on Peak Load Shaving in Microgrid—Potential Benefits, Challenges, and Future Trend. *Energies* 15 (6). https://doi.org/10.3390/en15062278.

Rana, Md M., Akhlaqur Rahman, Moslem Uddin, Md R. Sarkar, Sk A. Shezan, Md F. Ishraque, S. M. Rafin, and Mohamed Atef. 2022. A Comparative Analysis of Peak Load Shaving Strategies for Isolated Microgrid Using Actual Data. *Energies* 15 (1). https://doi.org/10.3390/en15010330.

Saboori, H., and H. Abdi. 2013. "Application of a grid scale energy storage system to reduce distribution network losses." 18th Electric Power Distribution Conference, 30 April-1 May 2013.

Schütz, Thomas, Rita Streblow, and Dirk Müller. 2015. "A comparison of thermal energy storage models for building energy system optimization." *Energy and Buildings* 93: 23-31. https://doi.org/https://doi.org/10.1016/j.enbuild.2015.02.031. https://www.sciencedirect.com/science/article/pii/S037877881500136X.

Sigrist, Lukas, Enrique Lobato, and Luis Rouco. 2013. "Energy storage systems providing primary reserve and peak shaving in small isolated power systems: An economic assessment." *International Journal of Electrical Power & Energy Systems* 53: 675-683. https://doi.org/https://doi.org/10.1016/j.ijepes.2013.05.046. https://www.sciencedirect.com/science/article/pii/S0142061513002524.

Uddin, Moslem, M. F. Romlie, M. F. Abdullah, ChiaKwang Tan, G. M. Shafiullah, and A. H. A. Bakar. 2020. "A novel peak shaving algorithm for islanded microgrid using battery energy storage system." *Energy* 196: 117084. https://doi.org/https://doi.org/10.1016/j.energy.2020.117084. https://www.sciencedirect.com/science/article/pii/S0360544220301912.

Uddin, Moslem, Mohd Fakhizan Romlie, Mohd Faris Abdullah, Syahirah Abd Halim, Ab Halim Abu Bakar, and Tan Chia Kwang. 2018. "A review on peak load shaving strategies." *Renewable and Sustainable Energy Reviews* 82: 3323-3332. https://doi.org/https://doi.org/10.1016/j.rser.2017.10.056. https://www.sciencedirect.com/science/article/pii/S1364032117314272.

Zhang, Liting, Yongwen Yang, Qifen Li, Weijun Gao, Fanyue Qian, and Lifei Song. 2021. "Economic optimization of microgrids based on peak shaving and $CO_2$ reduction effect: A case study in Japan." *Journal of Cleaner Production* 321: 128973. https://doi.org/https://doi.org/10.1016/j.jclepro.2021.128973. https://www.sciencedirect.com/science/article/pii/S0959652621031644.